\documentclass[usenatbib,useAMS]{mn2e}

\usepackage{float}
\usepackage{graphicx}
\usepackage{hyperref}
\newcommand{\mpc}{$h^{-1}$Mpc}

\title[Cosmic inhomogeneity and anomalous diffusion]{Characterising cosmic inhomogeneity with anomalous diffusion}
\author[D. Kraljic]{D. Kraljic$^1$ \\$^1$ Rudolf Peierls Centre for Theoretical Physics, University of Oxford, 1 Keble Road, Oxford OX1 3NP, United Kingdom\\}

\begin{document}

\maketitle

\label{firstpage}

\begin{abstract}
Dark matter (DM) clustering at the present epoch is investigated from a fractal viewpoint in order to determine the scale where the self-similar scaling property of the DM halo distribution transits to homogeneity. Methods based on well-established `counts-in-cells' as well as new methods based on anomalous diffusion and random walks are investigated. Both are applied to DM halos of the biggest N-Body simulation in the `Dark Sky Simulations' (DS) catalogue and an equivalent randomly distributed catalogue. Results based on the smaller `Millennium Run' (MR) simulation are revisited and improved. It is found that the MR simulation volume is too small and prone to bias to reliably identify the onset of homogeneity. Transition to homogeneity is defined when the fractal dimension of the clustered and random distributions cannot be distinguished within the associated uncertainties. The `counts-in-cells' method applied to the DS simulation then yields a homogeneity scale roughly consistent with previous work ($\sim 150$\,\mpc). The characteristic length-scale for anomalous diffusion to behave homogeneously is found to be at about 250\,\mpc. The behaviour of the fractal dimensions for a halo catalogue with the same two-point function as the original but with shuffled Fourier phases is investigated. The methods based on anomalous diffusion are shown to be sensitive to the phase information, whereas the `counts-in-cells' methods are not.

\end{abstract}


\section{Introduction}
The three dimensional distribution of matter observed on the sky exhibits a complex structure of nodes, filaments, walls, and voids. A similar ``Cosmic Web" structure appears in N-body DM simulations. Due to the biasing of the luminous matter with respect to the dominant DM component, the knowledge of the distribution of DM in the Universe comes mostly from such numerical simulations. This intricate distribution is far from the description used in standard models in cosmology which assumes a Friedmann-Lema\^itre-Robertson-Walker (FLRW) background with small density perturbations. FLRW assumption is based on the high degree of isotropy inferred from observations of the cosmic microwave background (CMB) (once the dipole is subtracted) and the `Copernican Cosmological Principle' \citep{2012CRPhy..13.%
.682C}. The inhomogeneities are assumed to be only at small scales, so they should become vanishingly small after averaging over large enough scales, such that the simple description applies. It is therefore important to identify the scale above which the statistical properties of the Universe do not depend on the location. This requires a definition of the notion of the transition to (statistical) homogeneity. 

One approach to characterise the complex structure in the universe is via fractal analysis. Fractal analysis can also be used to determine the transition to homogeneity in a statistical sense. The analysis has been applied to redshift galaxy surveys: \cite{Pan:2000yg} to PSCz; \cite{Hogg:2004vw} to SDSS LRG; \cite{Sarkar:2009iga} to SDSS DR6; \cite{Scrimgeour:2012wt} to WiggleZ; \cite{2012arXiv1212.4832C} to SDSS DR9) and to N-Body simulations (\cite{2010JCAP...03..006G},\cite{2012MNRAS.427.2613C}). The quantities of interest are the number of objects within a sphere of radius $r$, $n(<r)$, and its scaling with $r$. This is known as a `counts-in-cells' method. The scaling exponent (or fractal dimension) approaches 3 on the large scales, which is the case for a homogeneous distribution. The transition to homogeneity has been detected for scales of roughly 70\,\mpc\ all the way up to 180\,\mpc\ or not detected at all \citep{2009EL.....8649001S}. 

Many different definitions for the transition have been used. \cite{Yadav:2010cc} define the transition to homogeneity at the point where the clustered distribution scaling exponent cannot be distinguished from the homogeneous one within 1$\sigma$ uncertainty. The transition, of course, depends on the survey size and intrinsic clustering. They find, for the standard $\Lambda$CDM universe, the upper bound for the transition (according to their definition) to be about 260\,\mpc. 

Another definition for the transition has been developed by \cite{Scrimgeour:2012wt}. The definition is insensitive to the statistical error of the scaling exponent, but it arbitrarily chooses the transition to happen when the exponent reaches to 1\% from the homogeneous value. This enables comparing different surveys at the expense of arbitrariness. 

The `counts-in-cells' method was modified to test cosmic homogeneity with Shannon entropy \citep{2013MNRAS.430.3376P}. This made possible a discussion of the finite volume effect of the surveys and the effect of overlapping of the spheres used to measure the scaling dimension. Those effects contribute to the confinement bias (sample of spheres mainly coming from the centre of the survey with lots of overlap), which is especially important for inhomogeneous samples with large fluctuations.

Recent proposals to quantify the structure of the ``Cosmic Web'' focus on the topology, studying the scale-dependent Betti numbers \citep{2013arXiv1306.3640V} or the genus statistic \citep{2013arXiv1310.4278S}. The topological information can be used as a measure of cosmological parameters \citep{2010ApJ...715L.185P}.

This paper revisits and improves the results from \cite{2012MNRAS.427.2613C}, which uses the `counts-in-cells' method on `Millennium Run' (MR) simulation \citep{2005Natur.435..629S}. We use one of the biggest N-body simulations to date \citep{2012PDU.....1...50K} from the `Dark Sky Simulations' (DS) \citep{2014arXiv1407.2600S} to avoid the finite volume and overlapping bias, which appear in smaller simulations like the MR simulation when large distances are probed. It is found that MR-sized simulations or samples are inadequate to accurately measure the transition to homogeneity.

This paper extends the methods of fractal analysis to include anomalous diffusion on the DM halo distribution. Diffusion can be modelled as a continuum limit of random walkers on the underlying network \citep{1987PhR...150..263H}. It probes different properties of the underlying point set compared with the `counts-in-cells' method \citep{ben2000diffusion}. Anomalous diffusion exhibits a number of scaling exponents (or dimensions) that are different from the normal diffusion and is associated with a fractal structure of the network \citep{ben2000diffusion, sokol}. Thus, it can be used as another characterisation of the distribution of matter. The methods and relevant theoretical background are summarised in Section \ref{sec:Methods}. Section \ref{DS vs MR} summarises the Dark Sky Simulations. Results for the determination of the fractal exponents and their scale dependence for the `counts-in-cells' method and methods based on anomalous diffusion are presented in Section \ref{Results}.

\section{Methods and theoretical background} \label{sec:Methods}
Visual inspection of the distribution of matter in the universe exhibits a self-similar ``Cosmic Web" structure of filaments, walls, and voids. A point set (e.g. DM halos, galaxies) can be characterized via a number of scaling dimensions. Cosmologically, we are interested in the spatial scale above which the transition to a homogeneous distribution occurs, which is determined by scaling dimensions reaching the usual, homogeneous values. 

The simplest fractal dimension is the 'counts-in-cells' dimension,
\begin{equation}
n_i(<r)\propto r^{D_i}
\end{equation}
which counts the number of points within a sphere of radius $r$ centred on the point $i$ with the scaling fractal dimension $D_i$. An associated quantity, the correlation integral, is defined as in \citep{Bagla:2007tv,martinez2010statistics}:
\begin{equation}
C_2(r)=\frac{1}{NM}\sum_{i=1}^{M}n_i(<r)
\end{equation}
where $N$ is the total number of points in the distribution (formally $N \rightarrow \infty$) and $M$ is the number of centres on which spheres of radius $r$ have been positioned. This exhibits scaling:
\begin{equation}
C_2\propto r^{D_2}, \quad
D_2=\frac{\partial \log C_2(r)}{\partial \log r}
\end{equation}
The deviation of the correlation (or `counts-in-cells') dimension $D_2$ from the ambient value $D$ is due to the clustering, as it can be seen in the regime of weak clustering: $D_2(r) \simeq D - D(\bar{\xi}(r)-\xi(r))$ where $\xi(r)$ is the two-point correlation function and $\bar{\xi}(r)$ is its average up to radius $r$ \citep{Bagla:2007tv}.

The structure of matter in the Universe is highly irregular and does not exhibit just one simple scaling. Different moments of the correlation integral can be taken to extract more information from the distribution. The generalised correlation integral and dimension are:
\begin{equation}
C_q(r)=\frac{1}{NM}\sum_{i=1}^{M}n_i^{q-1}(<r), \quad
D_q=\frac{\partial \log C_q(r)}{\partial \log r}
\end{equation}
A mono-fractal has just one scaling $D_q=D_2$ at all scales, whereas a multi-fractal has a spectrum of $D_q$. Positive values of $q$ put more weight on the dense regions whereas negative values on the underdense ones. We can thus extract the scaling behaviour for points mainly located in clusters or voids.

\subsection{Transition to homogeneity}
The correlation dimension $D_2$ depends on the spatial scale $r$. We are interested in the distance above which the distribution of objects is homogeneous. In the framework of the fractal analysis, this happens when the scaling dimension is equal to the ambient dimension (e.g $D_2=3$). For finite samples, however, the ambient value cannot be reached exactly. Based on \cite{Yadav:2010cc}, a clustered distribution is defined to become homogeneous when $D^{clustered}_2$ cannot be distinguished from $D^{unclustered}_2$ of the unclustered distribution within their errors. That is, $D^{unclustered}_2-D^{clustered}_2=\Delta D_2  \simeq \sigma_{\Delta D_2}$, where $\sigma_{\Delta D_2}$ is the statistical error, which has been calculated to happen at about 260\,\mpc\, for the $\Lambda$CDM model. This definition is statistically well-motivated. However, this only considers the statistical contribution. Additional contributions to $\sigma_{\Delta D_2}$ due to cosmic variance, survey geometry and the selection function bias the above definition of the transition to homogeneity towards smaller values.

The unclustered\footnote{The terms ``unclustered'', ``random'', and ``homogeneous'' will be used interchangeably} distributions we use are of the same size as the clustered sample in terms of the volume and the number of objects, but in which the positions of the objects are random.

\subsection{Diffusion}
\subsubsection{Normal diffusion}
Diffusion on a point set or a network can be modelled with a large number of random walkers\footnote{For a short review see \cite{sokol}}. Diffusion is their continuum limit. The walkers jump to neighbouring points from their current position and thus traverse a certain distance, assuming the set of points is embedded in a space that has a notion of distance (e.g. Euclidean space). The standard result for the random walkers (assuming the uncorrelated nature of the walk and no directional bias) is:
\begin{eqnarray}
\langle \mathbf{R}^2 \rangle = \int P(r,t)\bmath r^2 d^d \bmath r \propto t \\
P(\bmath r,t|K,d)=\frac{1}{(4\pi K t)^{d/2}}\exp \left( -\frac{d \; r^2}{4Kt} \right)
\label{probeqn}
\end{eqnarray}
That is, the mean squared displacement $\langle \mathbf{R}^2 \rangle$ scales with the number of steps $t$ and is independent of the dimension $d$ of the Euclidean space. This is a property of normal diffusion. $P(\bmath r,t|K,d)$ is the probability distribution for the distance $\bmath r$ of the random walker from the origin after $t$ number of steps with $K$ being the diffusion coefficient and $d$ the dimensionality of the underlying space.

On a lattice point set the random walker jumps onto the neighbouring points. However, on non-lattice point sets random walkers have to be supplied with the notion of neighbouring points. To have a meaningful long-distance scaling, the prescription for the random walker on how to jump to the next point should be limited to short distances. The asymptotic scaling should also be robust against changes in the short-distance prescription. This can be achieved by linking each point to a fixed number of the nearest points, thus creating a network for the walkers. This prescription introduces a slight modification of the scaling law at short distances, while the asymptotic behaviour remains unchanged with respect to the number of nearest points the walker is allowed to jump on.

\subsubsection{Anomalous diffusion}
A diffusive process is said to be anomalous \citep{1987PhR...150..263H} if the mean squared displacement grows as:
\begin{equation}
\langle \langle \mathbf{R}_i^2 \rangle \rangle_i \propto t^\alpha 
\end{equation}
with $\alpha \neq 1$. Again, $t$ is the number of steps of the random walker, $i$ the initial position of the walker, and $\langle \cdot{} \rangle_i$ is the average over positions $i$. The values with $\alpha<1$ and $\alpha>1$ correspond to subdiffusion and superdiffusion.

The advantage of this method is that the anomalous diffusion exponent $\alpha$ is the same ($\alpha=1$) for homogeneous distributions in any dimension, unlike the `counts-in-cells' method which gives $D_2=1,2,3$ respectively for a homogeneous line, sheet, and volume distribution and does not necessarily imply fractal behaviour. For example, a galaxy located in a wall would have $D_2 \simeq 2$ but $\alpha \simeq 1$. Similarly as the 'counts-in-cells' method, the homogeneity is reached when $\alpha^{unclustered}(R)-\alpha^{clustered}(R)=\Delta \alpha(R) \simeq \sigma_{\Delta \alpha(R)}$ at a certain `radius' $R$. Note that the effective distance scale here is $\sqrt{\langle \bmath R^2 \rangle}$, meaning that the diffusion is sensitive to a range of scales.

Cosmological data (simulated or observed) is limited in volume. Therefore, a diffusion process must be stopped as it reaches the boundaries of the dataset to prevent unphysical leakage (or reflections) of random walkers from the volume. In other words, the whole probability distribution must be contained within the dataset volume to correctly evaluate the mean distance squared of the random walkers. This imposes an upper bound on the distances probed (smaller than characteristic length-scale of the sample) and makes this method suitable only for large-volume datasets.
\\
\\
Conversely one can consider:
\begin{equation}
\langle \langle T \rangle \rangle_i \propto R^{d_w}
\end{equation}
The walk dimension $d_W$ is defined via the mean number of steps $\langle T \rangle$ that a random walker needs in order to leave for the first time a ball of radius $R$ centred on the position $i$. This method has a clearly defined scale $R$, unlike the previously discussed method above that considers the mean square displacement $\sqrt{\langle \bmath R^2 \rangle}$. It is comparable to the `counts-in-cells' method in terms of efficiency of volume use. For normal diffusion we have $d_W=2$, and for a mono-fractal with one scaling exponent we have $d_W=2/\alpha$. Again, the distribution is homogeneous above the radius where $\Delta d_W (R) \simeq \sigma_{\Delta d_W (R)}$.

From Eqn.~\ref{probeqn} we see that the direct handle on the dimensionality of the underlying Euclidean or fractal space is the probability of return $P(\bmath r= \bmath 0,t)\propto t^{-d_s/2}$, where $d_s$ is the spectral dimension ($d_s=3$ for a homogeneous three dimensional distribution). However, this approach is difficult to apply on cosmological data that is limited in volume. Random walkers might leak out of the box and never return, thus biasing the probability of return. Also, spatial scale is not present directly. The scale can be indirectly inferred from the diffusion time $t$, i.e. long/short $t$ corresponds to large/small spatial scale.

\section{Dark Sky Simulations}
\label{DS vs MR}
Dark Sky (DS) Simulations \citep{2014arXiv1407.2600S} Early Data Release contains a large volume DM N-Body simulation using $10240^3 \approx 10^{12}$ particles in a volume of $(8000$\,\mpc$)^3$ with the $\Lambda$CDM cosmology ($\Omega_m=0.295,\Omega_b=0.0468,\Omega_\Lambda=0.705,n_s=0.969,h=0.688,\sigma_8=0.835$). The halo catalogue is obtained by a phase-space \textsc{rockstar} algorithm \citep{2013ApJ...762..109B} at $z=0$. For reasons of computational efficiency, we reduce the halo catalogue by employing a mass cut ($M_{vir}>10^{12}M_{\odot}/h$). The resulting catalogue contains $\approx 2.3 \cdot 10^9 $ DM halos. The mean number density is $4.6 \cdot 10^{-3}$ halos per (\mpc)$^3$. Equivalently, the mean inter-halo distance is about 6\,\mpc.

\section{Results and discussion}
\label{Results}
In all the determinations of the scaling exponents' dependence on the radial distance we use linear fits to the log-log data with 15\,\mpc\ bin size.

\subsection{`Counts-in-cells' dimension and the overlapping and confinement bias}
Here we revisit the results from \cite{2012MNRAS.427.2613C} that measured the `counts-in-cells' dimension $D_2(r)$ in the Millennium Run simulation. We address the shortcomings mostly stemming from the small size of the simulation (500\,\mpc\, box). The periodic boundary conditions used in the MR simulation limit the typical distance between two points to below 250\,\mpc\, and homogeneity is imposed automatically at this radius\footnote{Modes bigger than that are missing altogether}. This is at odds with the theoretical expectation in $\Lambda$CDM ($\sim 260$\,\mpc)\citep{Yadav:2010cc}. 

\cite{2012MNRAS.427.2613C} determined the `counts-in-cells' dimension by drawing a thousand randomly selected centres from the DM halo distribution that were at least a certain distance (`depth') away from the edges of the box (without taking advantage of the periodic boundary conditions). They noticed `spurious homogenisation' effects, specifically, $D_2(r)$ exceeding the ambient dimension $D_2=3$ at radii (`depths') larger than 160\,\mpc. The origin of this unphysical result is the fact that at larger radii the spheres were preferentially drawn from the central part of the simulation box and thus could not be considered statistically independent. Taking many large overlapping spheres, all with centres in a small central subregion of the simulation box makes the errors artificially low which in turn hides the underlying bias in the determination of $D_2$ (i.e. the `confinement' and `overlapping' bias \citep{2013MNRAS.430.3376P,2009EL.....8649001S}). Therefore, in order to overcome the shortcomings described above and reliably probe scales $\sim 100$\,\mpc\, a volume much bigger than the MR simulation is needed.

\begin{figure}
 \includegraphics[width=\columnwidth]{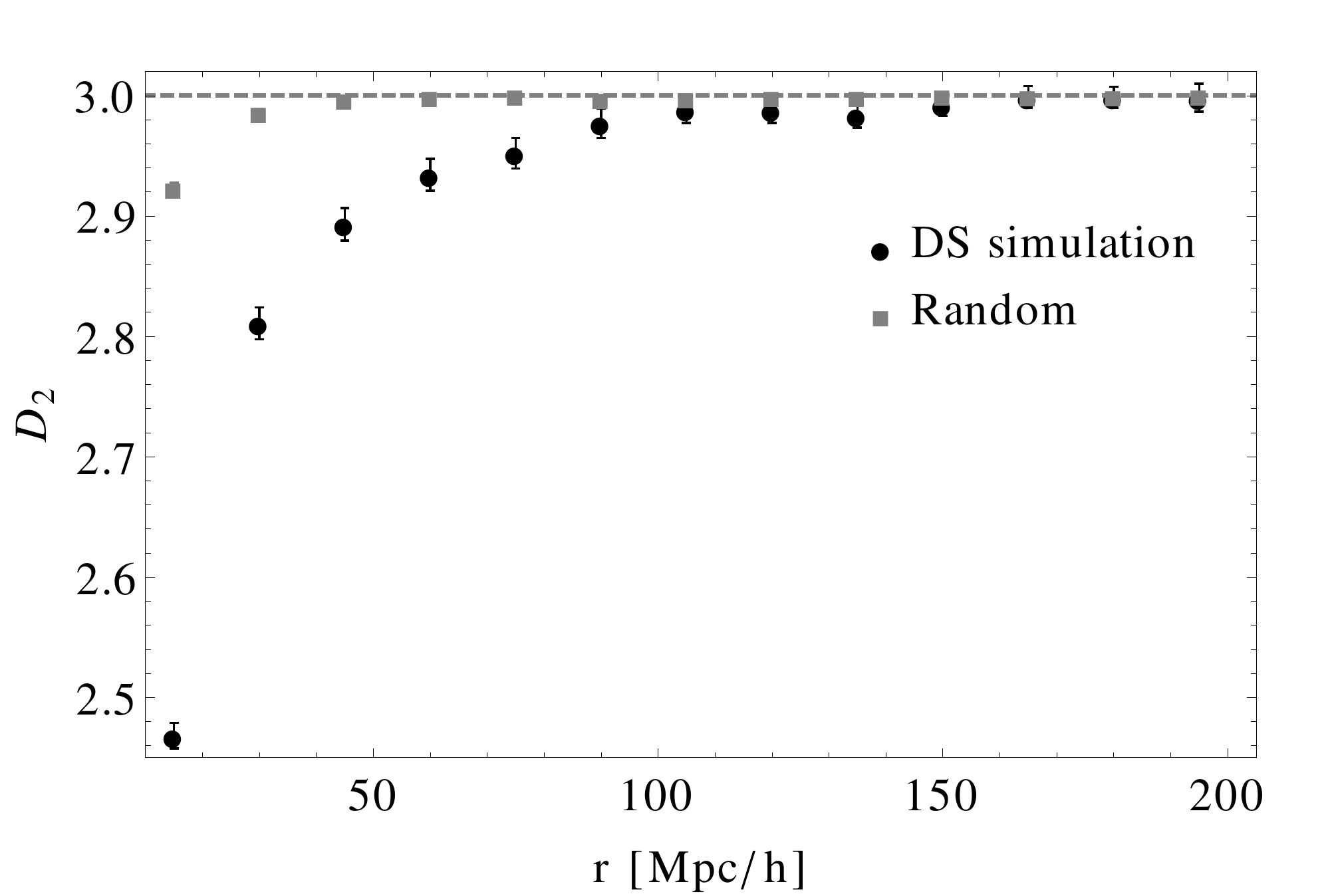}
 \caption{$D_2(r)$ for DS simulation and an equivalent random distribution. The scaling for the random distribution reaches $D_2=3$ very quickly. Note that the errorbars for the random distribution are much smaller compared to the clustered one.}
\label{spherescomp}
\end{figure}

\begin{figure}
 \includegraphics[width=\columnwidth]{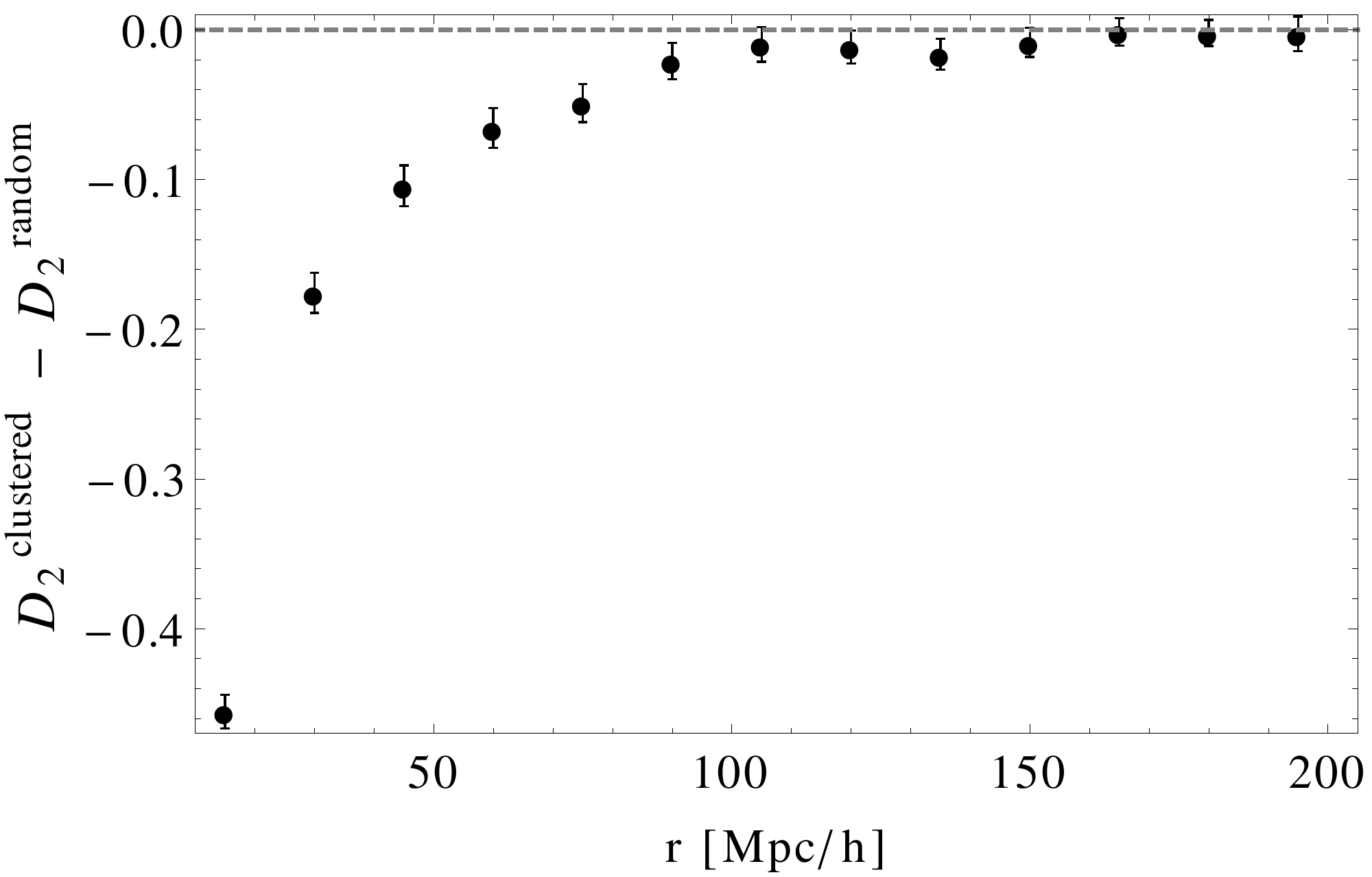}
 \caption{$\Delta D_2=D^{clustered}_2-D^{random}_2$ vs radius in \mpc\ for `counts-in-cells' method. The transition to homogeneity happens at around 150\,\mpc.}
\label{deltaspheres}
\end{figure}

The Dark Sky Simulation with a volume of $(8000$\,\mpc$)^3$ is big enough to draw a thousand spheres from it that do not overlap significantly. The correlation dimension $D_2(r)$ is calculated up to 200\,\mpc. The same method is applied to a random catalogue of the same size and the same number of points (see Fig.~\ref{spherescomp}). This compares to the work done before on the Millennium Run simulation DM halo catalogue, which used a thousand significantly \textit{overlapping} spheres drawn from it. 
\\
In Fig.~\ref{spherescomp} we see that the scaling for the random distribution reaches the expected value ($D_2(r) \simeq 3$) at about 40\,\mpc. Before that radius the effects of discreteness influence the scaling dimension (mean particle separation $\lambda$ is $\sim6$\,\mpc, so the scaling makes sense at radii $r\gg\lambda$). Hence, the relevant quantity that determines whether a clustered distribution can be distinguished from the random one is $\Delta D_2(r)=D^{clustered}_{2}(r) - D^{unclustered}_{2}(r)$, which is plotted in Fig.~\ref{deltaspheres}.

\begin{figure}
 \includegraphics[width=\columnwidth]{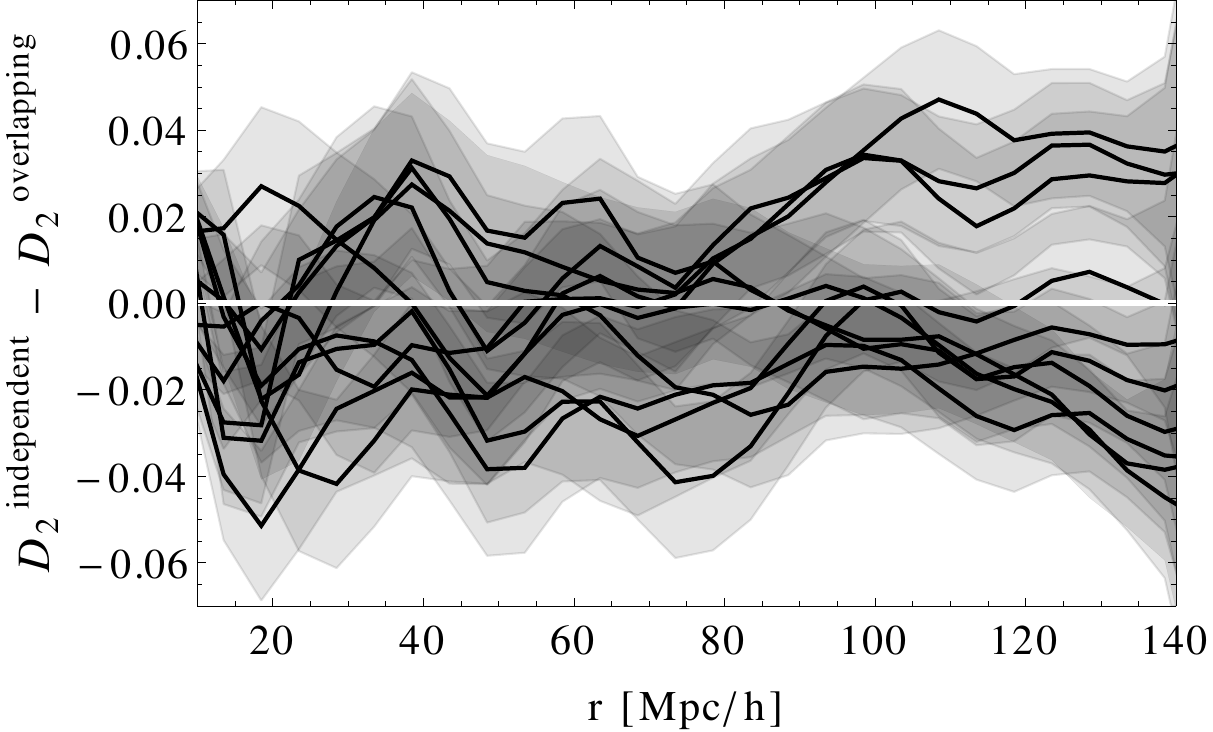}
\caption{Each black line represents the difference between the `counts-in-cells' dimension,  $\Delta D_2=D_2^{independent}-D_2^{overlapping}$, calculated in the full DS volume and in one of the ten ($500$\,\mpc)$^3$ subvolumes of DS where the spheres overlap substantially. `Depth' is 140\,\mpc. Shaded regions are errors ($\approx \pm 0.02$) corresponding to each line.}
\label{diffspheres}
\end{figure}

\begin{figure}
 \includegraphics[width=\columnwidth]{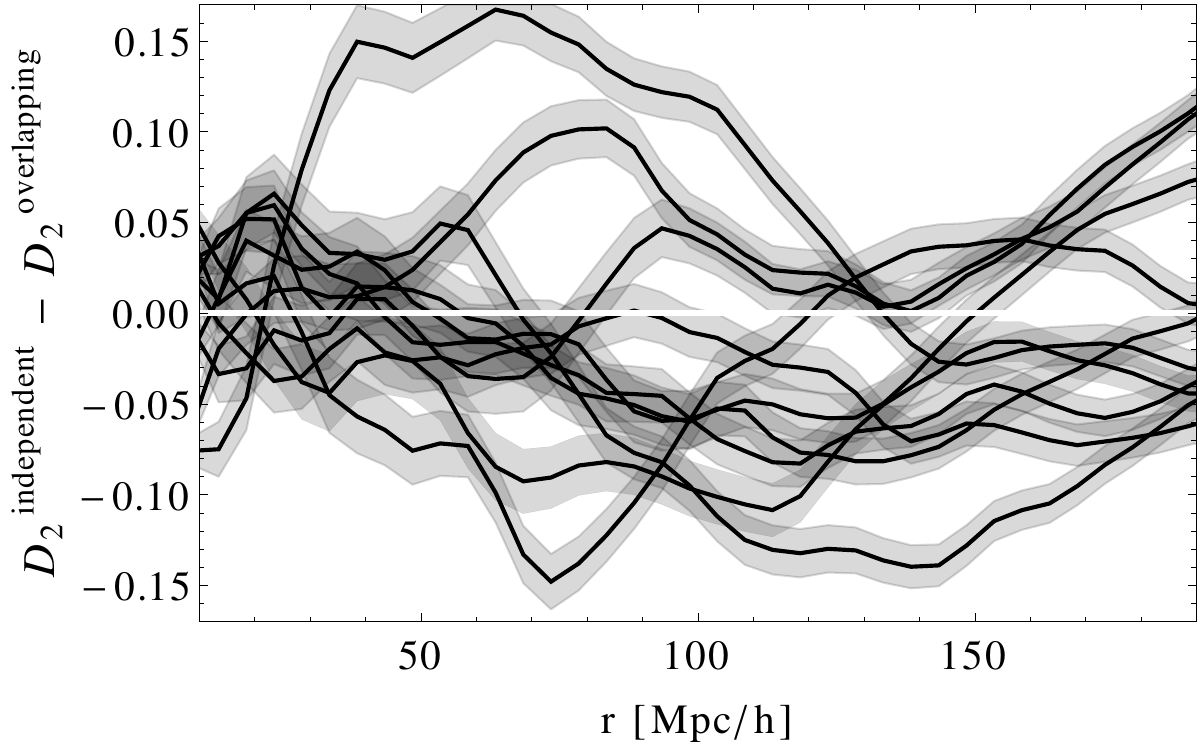}
 \caption{Each black line represents the difference between the `counts-in-cells' dimension,  $\Delta D_2=D_2^{independent}-D_2^{overlapping}$, calculated in the full DS volume and in one of the ten ($500$\,\mpc)$^3$ subvolumes of DS where the spheres overlap substantially. `Depth' is 200\,\mpc. Shaded regions are errors ($\approx \pm 0.02$) corresponding to each line.}
\label{diffspheres2}
\end{figure}

The `counts-in-cells' dimension $D_2(r)$ is always below $D_2=3$ all the way to 200\,\mpc, avoiding the spurious effects resulting from the significant overlap as discussed above. The transition to homogeneity happens at about 150\,\mpc\, which is higher but broadly comparable to previous work \citep{2012MNRAS.427.2613C}. 

We illustrate the `confinement' and `overlapping' bias by randomly selecting a thousand centres in each of 10 MR-sized $(500$\,\mpc$)^3$ subvolumes of the DS simulation. The exponent $D_2$ is determined for depths 200\,\mpc\ and 140\,\mpc\ in each subvolume. This is then compared to the $D_2$ exponent deduced by using a thousand centres from the full DS volume where the spheres are not overlapping significantly. We plot $\Delta D_2(r)=D_2^{independent}(r)-D_2^{overlapping}(r)$ in Figs.~\ref{diffspheres} and \ref{diffspheres2}. The bias is bigger and the errors relatively smaller for the depth of 200\,\mpc\ compared to 140\,\mpc\ since the sampled spheres overlap more and are more confined. The systematic offset persist across all scales, also the relevant ones at around 100\,\mpc. This suggests that volumes as small as the MR simulation are prone to systematic bias of the order of 1\%. Hence, the methods to determine the transition to homogeneity either by focussing on the statistical error in the scaling exponent, $\sigma_{\Delta D_2}$ \citep{Yadav:2010cc}, or proximity to the theoretical scaling exponent, $(D_2 - D)/D = 1\% $ \citep{Scrimgeour:2012wt}, cannot be applied reliably. This analysis also shows that the precise shape of the two-point correlation function, which is related to $D_2$ via $D_2(r) \simeq D - D(\bar{\xi}(r)-\xi(r))$, is also affected when measured in a MR size simulation or a survey. For a MR-sized simulation the biases can be mitigated by employing periodic boundary conditions which reduce confinement and overlapping of the spheres.

\subsection{Anomalous diffusion dimensions}

\begin{figure}
 \includegraphics[width=\columnwidth]{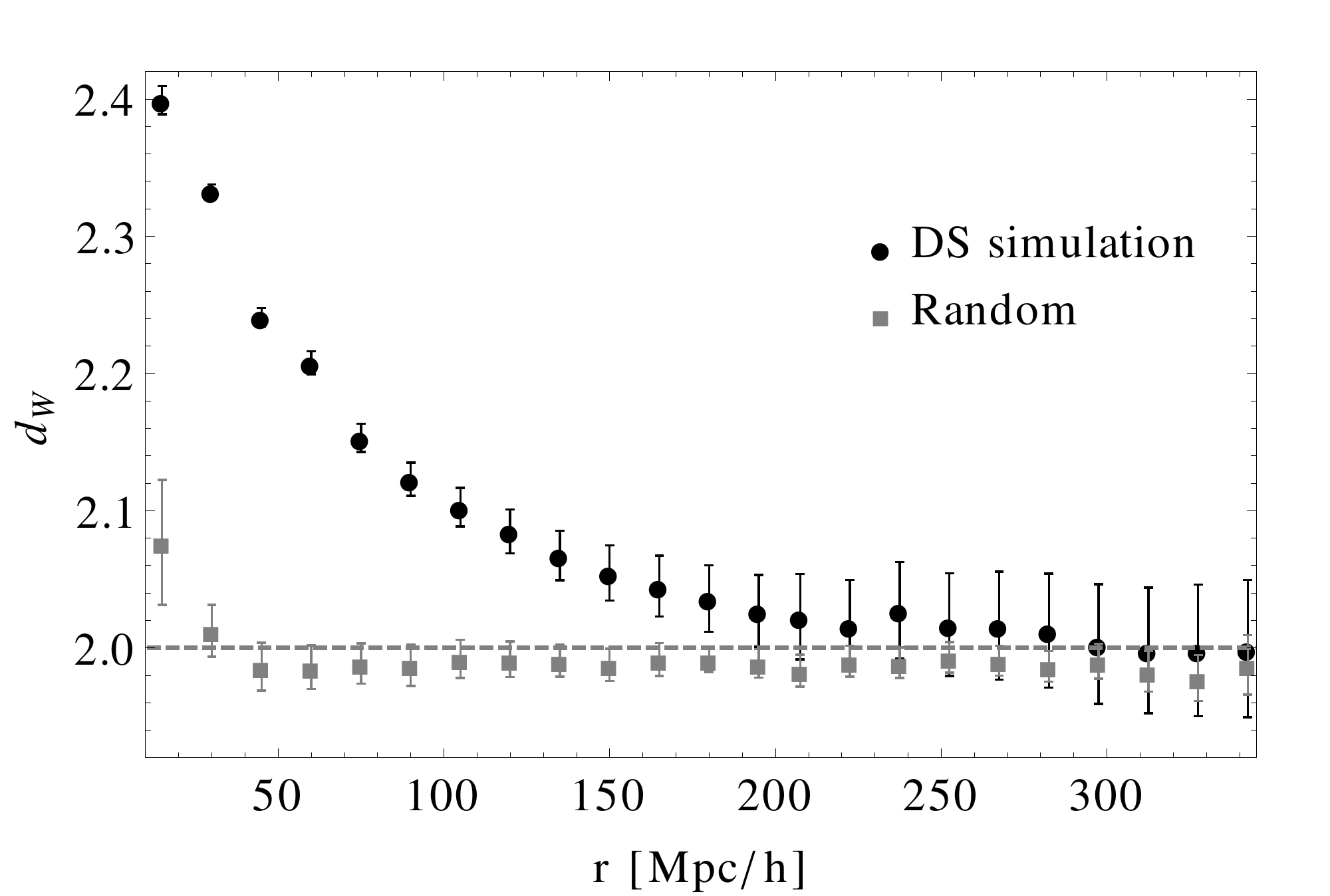}
 \caption{$d_W(r)$ for DS and equivalent implementation on a random catalogue. The small scale effects of the implementation of the random walkers is seen in the lower set of points (random catalogue). Anomalous diffusion persists beyond 200\,\mpc\ in the clustered distribution.}
\label{INVwalk}
\end{figure}

\begin{figure}
 \includegraphics[width=\columnwidth]{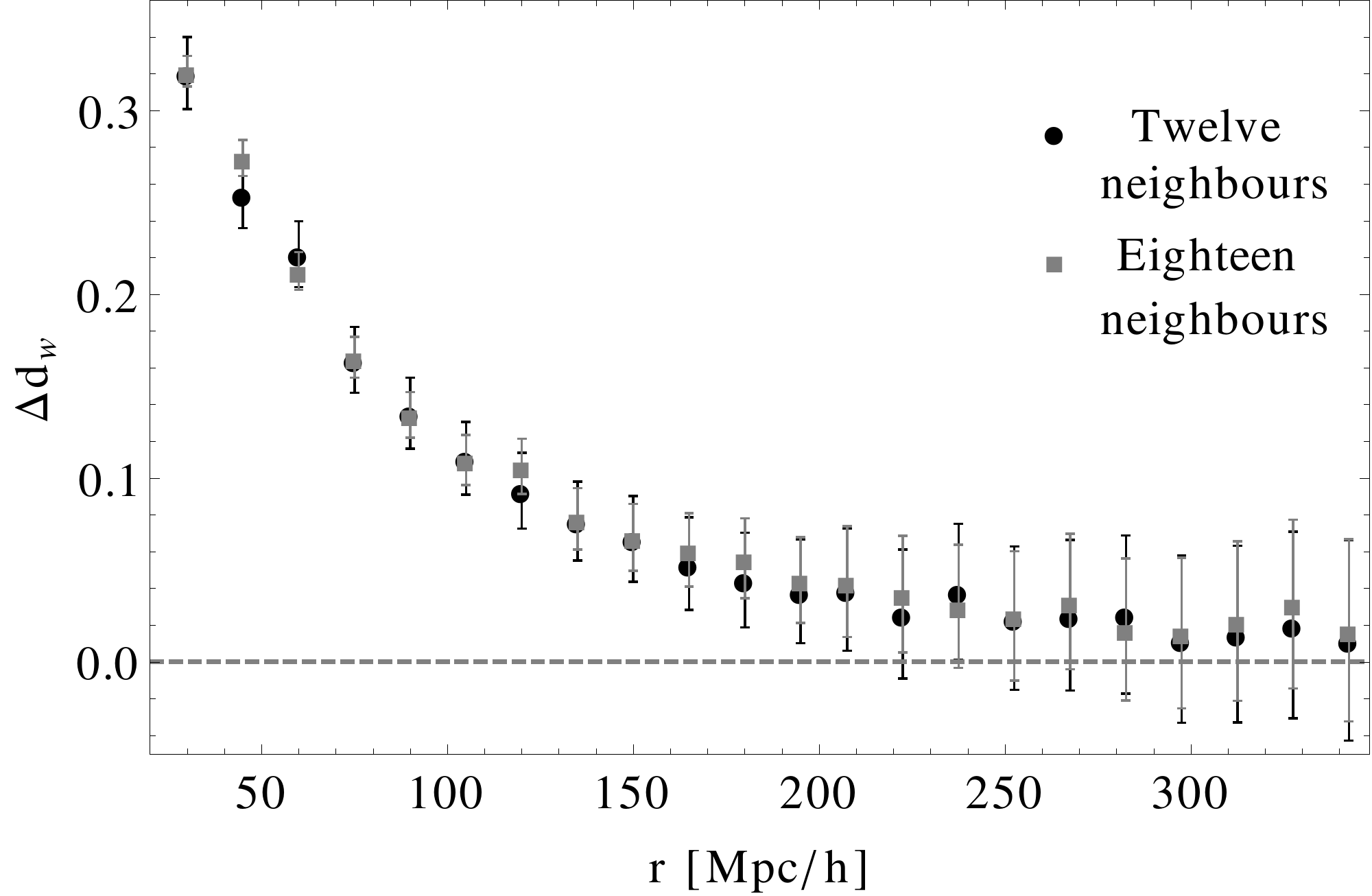}
 \caption{$\Delta d_W=d^{clustered}_W-d^{unclustered}_W$ between the DS sample and a random distribution for the implementation where the neighbourhood of each halo is the 12 or 18 nearest halos.}
\label{INVwalkdiff}
\end{figure}

\begin{figure}
 \includegraphics[width=\columnwidth]{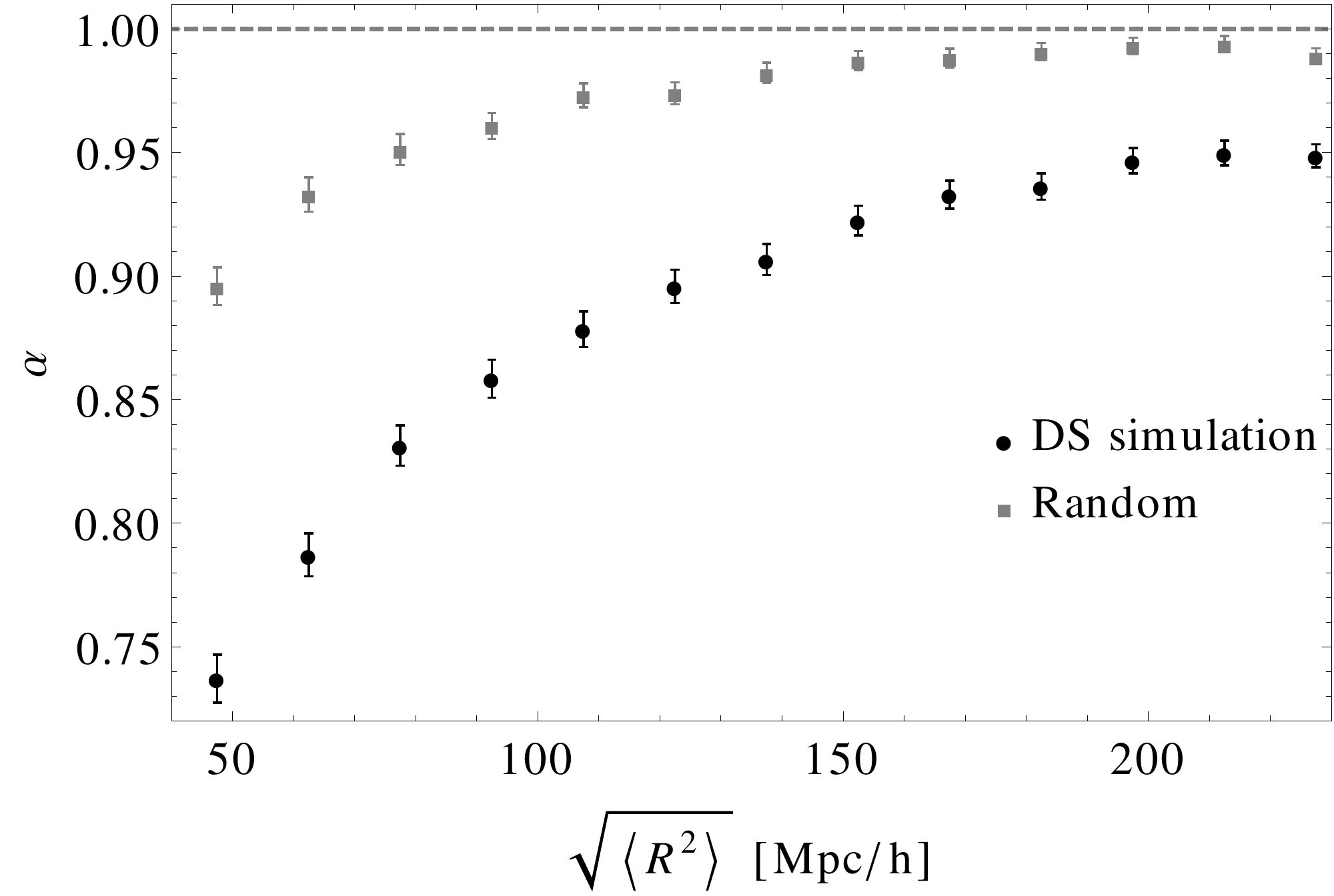}
 \caption{Mean square displacement scaling $\alpha(r)$ for a random distribution (upper set of points) and DS for the implementation where the neighbourhood of each halo is the nearest 12 halos.}
\label{walkalpha}
\end{figure}

\begin{figure}
 \includegraphics[width=\columnwidth]{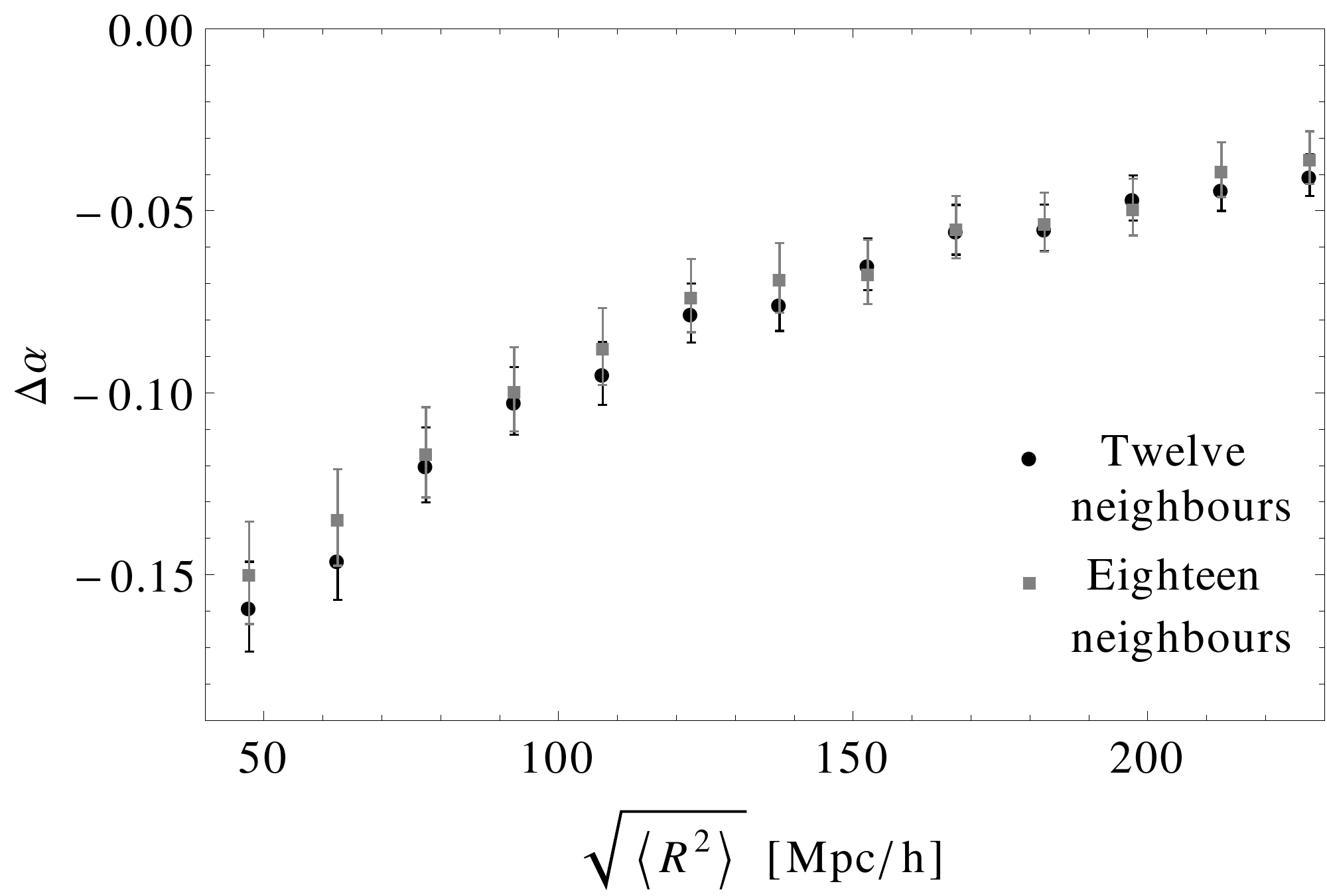}
 \caption{$\Delta \alpha=\alpha^{clustered} - \alpha^{unclustered}$ between the DS sample and a random distribution for the implementation where the neighbourhood of each halo  is the nearest 12 or 18 halos.}
\label{walkdeltaalpha18}
\end{figure}

In this section we discuss a new method of characterising the cosmic structure as seen in DM halos in N-Body simulations. It is based on the concept of anomalous diffusion (Sec. \ref{sec:Methods}) which we model using random walkers. 

The network used for random walkers was constructed in such a way that each DM halo had a neighbourhood of 12 nearest halos. The construction of the network is computationally expensive, especially for large catalogues. Additionally, the links that were farther than 20\,\mpc\ were rejected in order to keep short-distance effects from influencing the long-distance scaling. This lower `cut-off' provides a separation of small scales from the large ones that are of interest here. However, together with the discreteness of the data, the `cut-off' introduces spurious effects at smaller scales which can be seen for the diffusion on the unclustered random distribution of halos (see Fig. \ref{INVwalk} and Fig. \ref{walkalpha}). Note how the scaling for the random distribution approaches the expected theoretical values. To correct for this spurious effect, the difference of scaling exponents between the clustered and unclustered distributions is taken as relevant for the study of the transition to homogeneity. 

The analysis was repeated for a network constructed by connecting each halo to the 18 nearest halos. This was done in order to check that the results for the scaling dimensions did not substantially depend on such a choice (see Fig.~\ref{INVwalkdiff} and Fig.~\ref{walkdeltaalpha18}). Some differences are expected at smaller scales due to spurious discreetness effects.

The method was applied to a set of lattice distributions and deterministic fractal distributions with known scaling exponents in order to confirm that the implementation of the diffusive process gave correct asymptotic scaling exponents.

The number of walkers for each starting point is chosen to be very large in order for the walkers to saturate the possible ways of exploring a given volume starting from a specific point. Thus, the spread in the scaling exponents comes dominantly from having different points used as centres for the random walkers.

\subsubsection{Mean square displacement scaling $\alpha$}
Scaling of the mean distance squared with the number of steps, $\langle \bmath R^2 \rangle \propto t^\alpha$, up to an effective radius $\sqrt{\langle \bmath R^2 \rangle}$ of 220\,\mpc\, was explored. The number of starting centres for the random walkers was 512 and care was taken to make sure that the diffusion processes were terminated before overlapping significantly. 

This is a less efficient method in terms of using the available volume compared to the 'counts-in-cells' method. The reason is that, in order to evaluate the scaling exponent $\alpha$, all the random walkers for a given number of steps $t_0$ are needed. Hence, the diffusive process extends over a range of scales (with the characteristic scale $\sqrt{\langle \bmath R^2(t_0) \rangle}$ much smaller than $ |\bmath R_{max}(t_0)|$). As a result, the transition to homogeneity happens slowly as the notion of scale is spread out (see Figs.~\ref{walkalpha} and \ref{walkdeltaalpha18}). This explains the slow convergence of $\alpha$ for the random distribution to the theoretical value $\alpha=1$ as well as the slow convergence of the clustered distribution to the random one up to $\sqrt{\langle \bmath R^2 \rangle} \sim 220$\,\mpc.

\subsubsection{Walk dimension $d_W$}
Scaling of the mean number of steps needed to leave a ball of radius $R$, $\langle T \rangle \propto R^{d_w}$ , was determined up to 350\,\mpc. The same volume of $(8.0h^{-1}$\,Gpc$)^3$ as in the `counts-in-cells' method was used (same efficiency of volume use). This enables a direct comparison of the results and the errorbars. The distance scale for this diffusive process is well-defined, contrary to the method measuring $\alpha$ discussed above. This can be seen from the walk dimension $d_W$ and anomalous diffusion dimension $\alpha$ for the unclustered distribution in Figs.~\ref{INVwalk} and \ref{walkalpha}. The walk dimension, $d_W$, converges to the theoretically expected value at about 50\,\mpc\ whereas $\alpha$ takes longer. The transition to homogeneity via anomalous diffusion, as defined in Sec.~\ref{sec:Methods}, happens at a different, larger scale (above 250\,\mpc) than for the `counts-in-cells' method (see Figs.~\ref{INVwalk} and ~\ref{INVwalkdiff}).

\begin{figure}
 \includegraphics[width=\columnwidth]{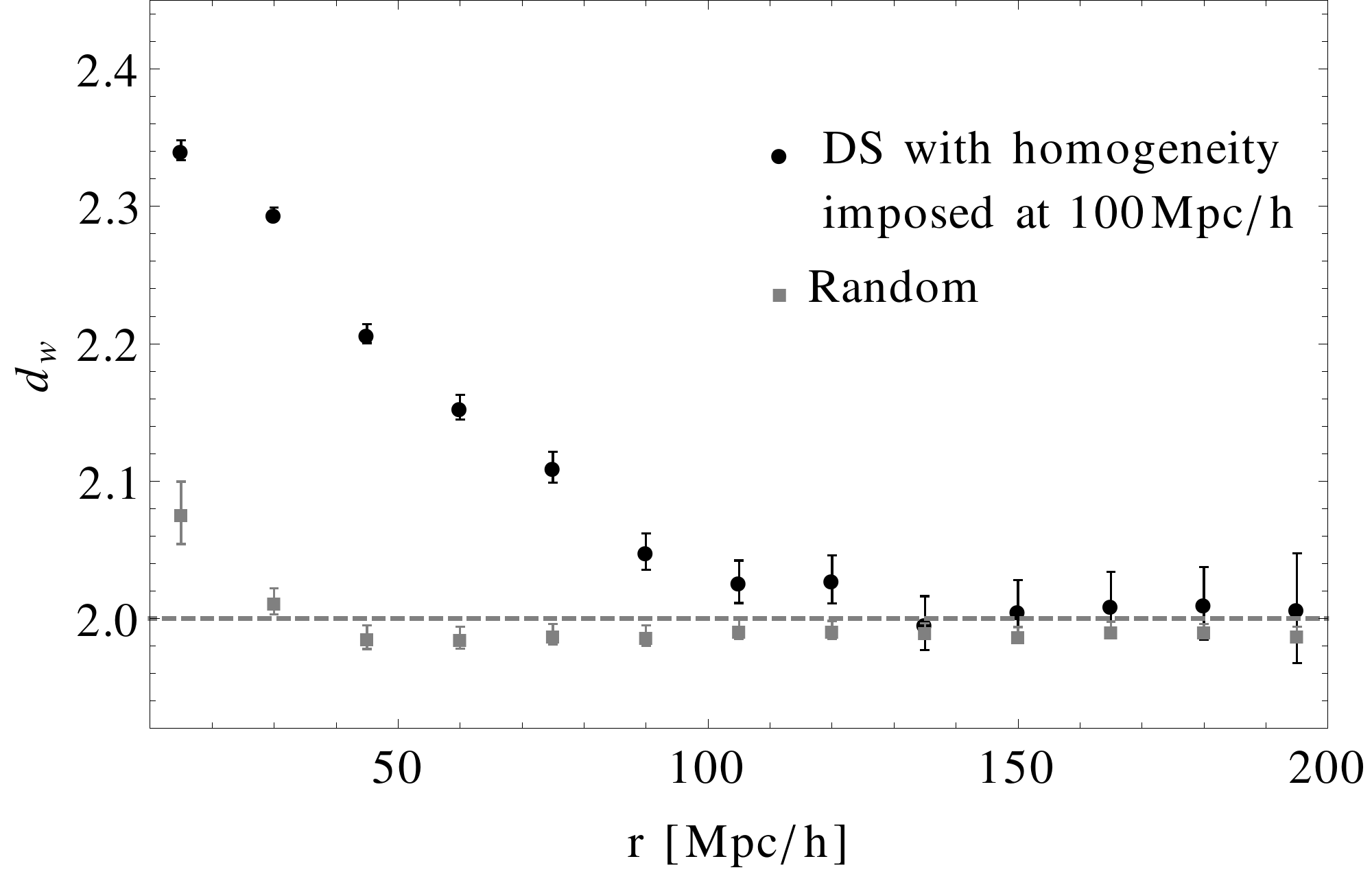}
 \caption{$d_W(r)$ for the case where the homogeneity is artificially imposed at 100\,\mpc. The lower set of points corresponds to the unclustered distribution.}
\label{INVhom}
\end{figure}

It is useful to check how sensitive the walk dimension $d_W$ is to the transition to homogeneity. This is done by randomly sampling subvolumes of the DS simulation with 100\,\mpc\ box lengths and imposing periodic boundary conditions to achieve homogeneity at this scale. In Fig.~\ref{INVhom} we see that the transition to homogeneity does begin at about 100\,\mpc. This should be compared with Fig.~\ref{INVwalk}, where the transition happens later, as described above.

\subsection{Fractal analysis and phase information}
The `counts-in-cells' method has been used to test the assumption of homogeneity and determine the transition to it. The method, despite being inspired by fractal analysis, cannot characterise the intricate filamentary structure of the ``Cosmic Web''. The reason is that the `counts-in-cells' method, provided the homogeneity is reached in the sample, only contains the information about the two-point correlation function or equivalently the power spectrum \citep{Bagla:2007tv}. This means that the phase information of the density field which is responsible for the filamentary structures is lost. 

Following \cite{2001MNRAS.325..405C} and \cite{2003eucm.book..237C}, the phases of the Fourier transform of the density field obtained from the DS halo catalogue, $\delta_{\mathbf k}$, were shuffled. This erases the phase information but preserves the power spectrum that depends only on the square of the density field $|\delta_{\mathbf k}|^2$. The density field with shuffled phases was then used to generate a discrete distribution of points (i.e. a new halo catalogue). This was done by Poisson sampling from the phase-shuffled density field. That is, in a volume $V$, the number of selected points is proportional to the enclosed mass $\langle N(V)\rangle \propto M(V)$. The total number of points selected is such that the same overall number density as in the DS catalogue is achieved. 

\begin{figure}
 \includegraphics[width=\columnwidth]{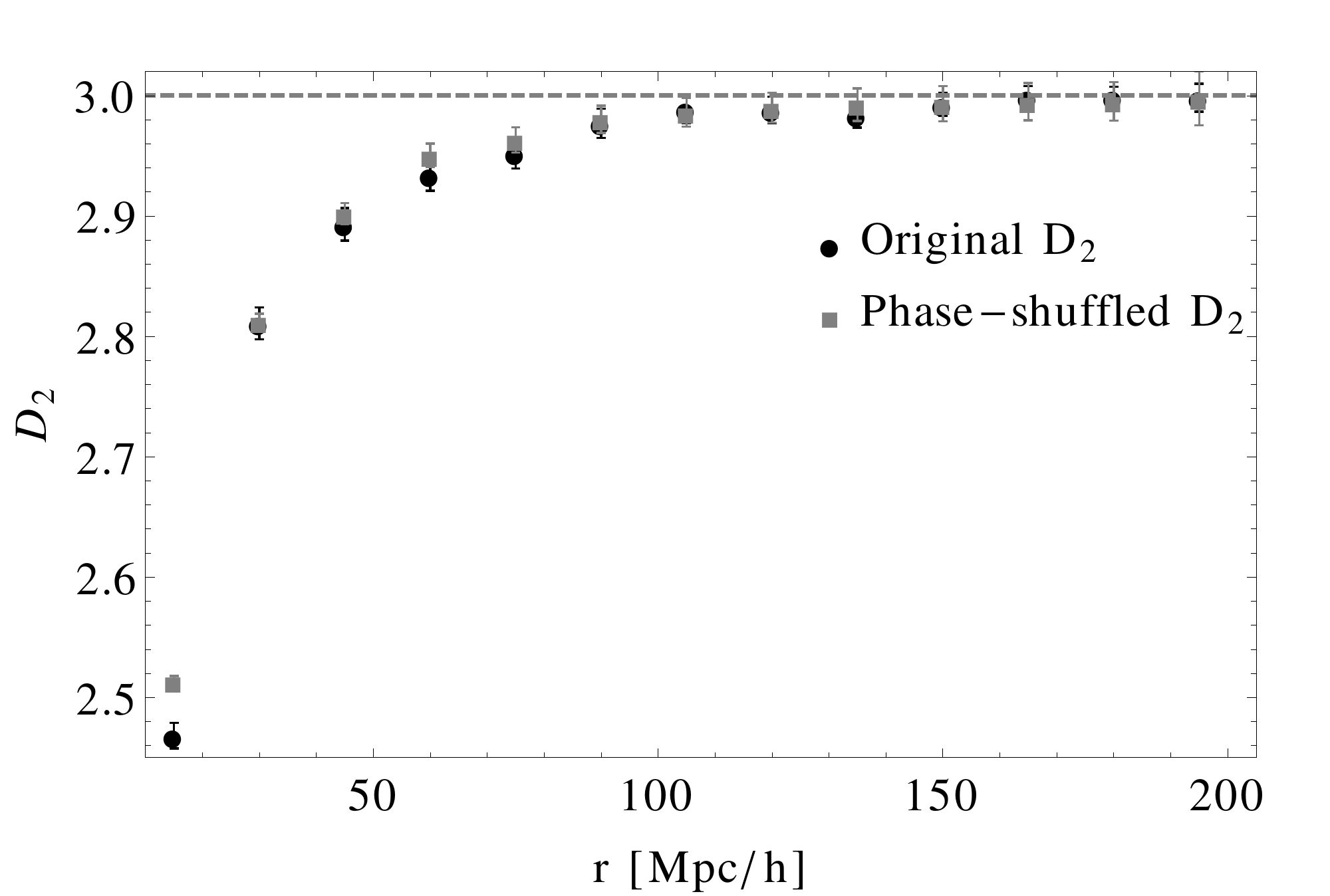}
 \caption{$D_2(r)$ for DS simulation and a catalogue with shuffled phases. The two distributions cannot be differentiated by looking at the `counts-in-cells' dimension $D_2(r)$.}
\label{shuf1}
\end{figure}

\begin{figure}
 \includegraphics[width=\columnwidth]{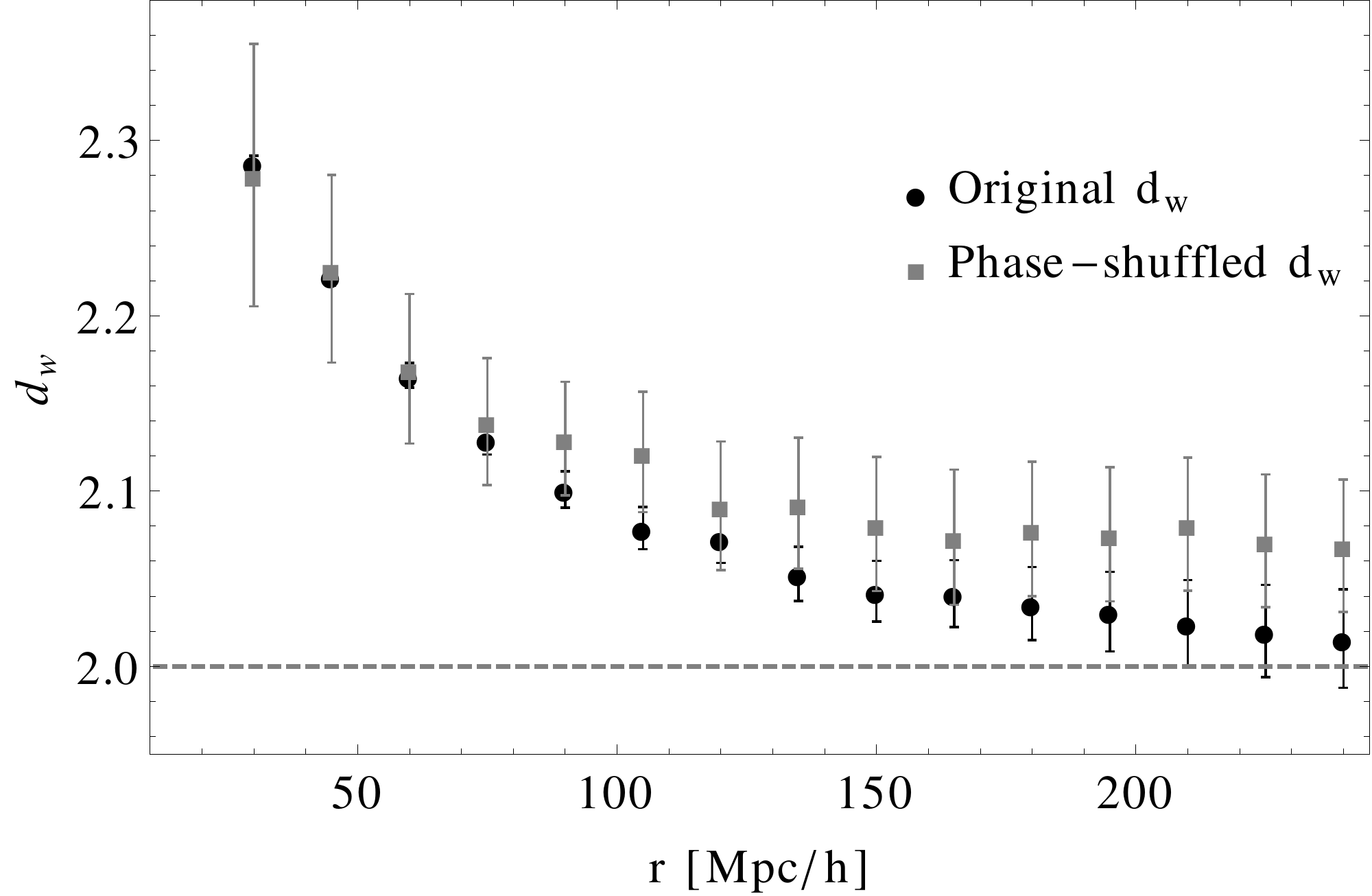}
 \caption{$d_w(r)$ for DS simulation and a catalogue with shuffled phases. At larger radii there is a difference in the walk dimension between the two distributions.}
\label{shuf2}
\end{figure}

The $D_2(r)$ determined in the generated catalogue is the same (within the errors) as in the original DS halo catalogue (Fig.~\ref{shuf1}). The changes in the scaling exponents for the diffusive processes can be checked as well. In Fig.~\ref{shuf2}, we see that the difference between the $d_W(r)$ for the original and the shuffled distribution shows systematic deviation above 100\,\mpc. We neglect smaller scales, because there the small scale effects coming from the process of obtaining a discrete representation of the phase-shuffled density field are the biggest and the scaling exponent $d_W(r)$ has little meaning.

\section{Conclusions}
This paper presents a novel way of characterising the distribution of matter in the universe based on the methods of anomalous diffusion. The methods were tested on the biggest DM halo catalogue from the `Dark Sky Simulations' (DS), a very large N-body simulation of $(8h^{-1}$\,Gpc$)^3$ volume. Diffusion was modelled by performing random walks on the set of halos. The underlying distribution was described by scaling exponents for various diffusive processes. Determination of the values of the scaling exponents and their dependence on the distance scale was used to decide whether a particular distribution could be distinguished from the homogeneous one.

We measure the mean number of steps $\langle T \rangle$ for a random walker to leave a ball of radius $R$, and how it scales with the radius: $\langle T \rangle \propto R^{d_w}$. In terms of efficiency of volume use, this method is comparable to the `counts-in-cells' method, which has been extensively used in the past to characterise the distribution of matter in the universe. For an explicitly homogeneous distribution (i.e. random, unclustered) we find quick convergence of $d_W$ to the theoretically expected value, namely $d_W\simeq2$. For the diffusion on the DS distribution of halos we find the convergence of the scaling exponent $d_W$ to the homogeneous value to be at about 250\,\mpc. This characteristic distance can then be measured and compared in different surveys and N-body simulations.

We also measure how the RMS distance traversed by a random walker scales with the number of steps, $\langle \bmath R^2 \rangle \propto t^\alpha$. The notion of distance here is only effective, $\sqrt{\langle \bmath R^2 \rangle}$, and therefore sensitive to a range of scales. Unsurprisingly the convergence of the scaling exponent $\alpha$ to the homogeneous values is slow. Measuring $\alpha$ probes shorter distance scales compared to the method above in a finite volume sample and is less efficient in the use of the available volume.

The size of the DS simulation also enables the improvement in the determination of the `counts-in-cells' fractal dimension compared to previous work that used smaller volumes. We show that Millennium Run sized volumes used in past studies of the `counts-in-cells' fractal dimension are not sufficiently big to precisely and accurately determine the transition to homogeneity, irrespective of the exact definition when transition happens. In particular, not taking advantage of the periodic boundary conditions resulted in preferential sampling of the central regions and consequently led to `confinement' bias. The substatial overlapping of the spheres also meant artificial reduction of the errorbars. Our work finds that the transition to homogeneity, as measured by the `counts-in-cells' dimension, happens at around 150\,\mpc\ which is slightly higher but broadly consistent with previous studies.

We investigate the behaviour of the fractal dimensions for a halo catalogue with the same two-point function as the original catalogue but with the Fourier phases shuffled. The methods based on anomalous diffusion can differentiate between the two cases, whereas the `counts-in-cells' method cannot.

\section*{Acknowledgements}
The author is supported by the STFC and would like to thank Subir Sarkar for useful discussions and comments on the draft.

\bibliography{refs.bib}

\begin{thebibliography}{}

\bibitem[\protect\citeauthoryear{Bagla, Yadav \& Seshadri}{Bagla
  et~al.}{2007}]{Bagla:2007tv}
Bagla J.,  Yadav J.,    Seshadri T.,  2007, Mon.Not.Roy.Astron.Soc., 390, 829

\bibitem[\protect\citeauthoryear{{Behroozi}, {Wechsler} \& {Wu}}{{Behroozi}
  et~al.}{2013}]{2013ApJ...762..109B}
{Behroozi} P.~S.,  {Wechsler} R.~H.,    {Wu} H.-Y.,  2013, \apj, 762, 109

\bibitem[\protect\citeauthoryear{Ben-Avraham \& Havlin}{Ben-Avraham \&
  Havlin}{2000}]{ben2000diffusion}
Ben-Avraham D.,  Havlin S.,  2000, Diffusion and reactions in fractals and
  disordered systems.
Cambridge University Press

\bibitem[\protect\citeauthoryear{{Chac{\'o}n-Cardona} \&
  {Casas-Miranda}}{{Chac{\'o}n-Cardona} \&
  {Casas-Miranda}}{2012a}]{2012MNRAS.427.2613C}
{Chac{\'o}n-Cardona} C.~A.,  {Casas-Miranda} R.~A.,  2012a, \mnras, 427, 2613

\bibitem[\protect\citeauthoryear{{Chac{\'o}n-Cardona} \&
  {Casas-Miranda}}{{Chac{\'o}n-Cardona} \&
  {Casas-Miranda}}{2012b}]{2012arXiv1212.4832C}
{Chac{\'o}n-Cardona} C.~A.,  {Casas-Miranda} R.~A.,  2012b, preprint
  (arXiv:1212.4832)

\bibitem[\protect\citeauthoryear{{Chiang}}{{Chiang}}{2001}]{2001MNRAS.325..405C}
{Chiang} L.-Y.,  2001, \mnras, 325, 405

\bibitem[\protect\citeauthoryear{{Clarkson}}{{Clarkson}}{2012}]{2012CRPhy..13..682C}
{Clarkson} C.,  2012, Comptes Rendus Physique, 13, 682

\bibitem[\protect\citeauthoryear{{Coles}}{{Coles}}{2003}]{2003eucm.book..237C}
{Coles} P.,  2003, {Statistical Properties of Cosmological Fluctuations}.
Kluwer Academic Publishers, p.~237

\bibitem[\protect\citeauthoryear{{Gaite}}{{Gaite}}{2010}]{2010JCAP...03..006G}
{Gaite} J.,  2010, \jcap, 3, 6

\bibitem[\protect\citeauthoryear{{Haus} \& {Kehr}}{{Haus} \&
  {Kehr}}{1987}]{1987PhR...150..263H}
{Haus} J.~W.,  {Kehr} K.~W.,  1987, \physrep, 150, 263

\bibitem[\protect\citeauthoryear{Hogg, Eisenstein, Blanton, Bahcall, Brinkmann
  et~al.,}{Hogg et~al.}{2005}]{Hogg:2004vw}
Hogg D.~W.,  Eisenstein D.~J.,  Blanton M.~R.,  Bahcall N.~A.,  Brinkmann J.,
   et~al., 2005, Astrophys.J., 624, 54

\bibitem[\protect\citeauthoryear{{Kuhlen}, {Vogelsberger} \& {Angulo}}{{Kuhlen}
  et~al.}{2012}]{2012PDU.....1...50K}
{Kuhlen} M.,  {Vogelsberger} M.,    {Angulo} R.,  2012, Physics of the Dark
  Universe, 1, 50

\bibitem[\protect\citeauthoryear{Martinez \& Saar}{Martinez \&
  Saar}{2010}]{martinez2010statistics}
Martinez V.~J.,  Saar E.,  2010, Statistics of the galaxy distribution.
CRC Press

\bibitem[\protect\citeauthoryear{Pan \& Coles}{Pan \& Coles}{2000}]{Pan:2000yg}
Pan J.,  Coles P.,  2000, Mon.Not.Roy.Astron.Soc., 318, L51

\bibitem[\protect\citeauthoryear{{Pandey}}{{Pandey}}{2013}]{2013MNRAS.430.3376P}
{Pandey} B.,  2013, \mnras, 430, 3376

\bibitem[\protect\citeauthoryear{{Park} \& {Kim}}{{Park} \&
  {Kim}}{2010}]{2010ApJ...715L.185P}
{Park} C.,  {Kim} Y.-R.,  2010, \apjl, 715, L185

\bibitem[\protect\citeauthoryear{{Sarkar}, {Yadav}, {Pandey} \&
  {Bharadwaj}}{{Sarkar} et~al.}{2009}]{Sarkar:2009iga}
{Sarkar} P.,  {Yadav} J.,  {Pandey} B.,    {Bharadwaj} S.,  2009, \mnras, 399,
  L128

\bibitem[\protect\citeauthoryear{Scrimgeour, Davis, Blake, James, Poole
  et~al.,}{Scrimgeour et~al.}{2012}]{Scrimgeour:2012wt}
Scrimgeour M.,  Davis T.,  Blake C.,  James J.~B.,  Poole G.,    et~al., 2012,
  Mon.Not.Roy.Astron.Soc., 425, 116

\bibitem[\protect\citeauthoryear{{Skillman}, {Warren}, {Turk}, {Wechsler},
  {Holz} \& {Sutter}}{{Skillman} et~al.}{2014}]{2014arXiv1407.2600S}
{Skillman} S.~W.,  {Warren} M.~S.,  {Turk} M.~J.,  {Wechsler} R.~H.,  {Holz}
  D.~E.,    {Sutter} P.~M.,  2014, preprint (arXiv:1407.2600)

\bibitem[\protect\citeauthoryear{Sokolov}{Sokolov}{2011}]{sokol}
Sokolov I.,  2011, in Meyers R.~A.,  ed., , Mathematics of Complexity
  and Dynamical Systems.
Springer New York, pp 13--25

\bibitem[\protect\citeauthoryear{{Speare}, {Gott}, {Kim} \& {Park}}{{Speare}
  et~al.}{2015}]{2013arXiv1310.4278S}
{Speare} R.,  {Gott} J.~R.,  {Kim} J.,    {Park} C.,  2015, \apj, 799, 176

\bibitem[\protect\citeauthoryear{{Springel} \& {et al.}}{{Springel} \& {et
  al.}}{2005}]{2005Natur.435..629S}
{Springel} V.,  {et al.} 2005, \nat, 435, 629

\bibitem[\protect\citeauthoryear{{Sylos Labini}, {Vasilyev}, {Pietronero} \&
  {Baryshev}}{{Sylos Labini} et~al.}{2009}]{2009EL.....8649001S}
{Sylos Labini} F.,  {Vasilyev} N.~L.,  {Pietronero} L.,    {Baryshev} Y.~V.,
  2009, EPL (Europhysics Letters), 86, 49001

\bibitem[\protect\citeauthoryear{{van de Weygaert} \& {et al.}}{{van de
  Weygaert} \& {et al.}}{2013}]{2013arXiv1306.3640V}
{van de Weygaert} R.,  {et al.} 2013, Lecture Notes in Computer Science. Vol.
  6970, Transactions on Computational Science XIV, pp 60--110

\bibitem[\protect\citeauthoryear{Yadav, Bagla \& Khandai}{Yadav
  et~al.}{2010}]{Yadav:2010cc}
Yadav J.~K.,  Bagla J.,    Khandai N.,  2010, Mon.Not.Roy.Astron.Soc., 405,
  2009

\end{thebibliography}
\bibliographystyle{mn2e}

\end{document}